\documentclass[aps,prb,reprint,floatfix,citeautoscript,longbibliography,superscriptaddress]{revtex4-1}
\usepackage{graphicx}
\usepackage{bm,amsmath,amssymb,mathrsfs,dcolumn}
\usepackage[usenames]{color}
\usepackage[colorlinks=true, linkcolor=blue, citecolor=blue, urlcolor=blue, linktoc=page, bookmarks=false, pdfstartview={FitH}, pdfborder={0 0 0.0 [3 3]}]{hyperref} 
\hypersetup{pdfborder=0 0 0,colorlinks=true,citecolor=blue,linkcolor=blue}
\usepackage[final]{pdfpages}
\usepackage{subfigure}
\usepackage{ulem}
\begin{document}
\title{Influence of nonlocal damping on the field-driven domain wall motion}

\author{H. Y. Yuan}
\author{Zhe Yuan}
\email[Corresponding author: ]{zyuan@bnu.edu.cn}
\author{Ke Xia}
\affiliation{The Center for Advanced Quantum Studies and Department of Physics,
Beijing Normal University, Beijing 100875, China}
\author{X. R. Wang}
\affiliation{Department of Physics, The Hong Kong University of
Science and Technology, Clear Water Bay, Kowloon, Hong Kong}
\affiliation{HKUST Shenzhen Research Institute, Shenzhen 518057, China}
\date{\today}

\begin{abstract}
We derive the complete expression of nonlocal damping in noncollinear 
magnetization due to the nonuniform spin current pumped by precessional magnetization 
and incorporate it into a generalized Thiele equation to study its effects on the dynamics 
of the transverse and vortex domain walls (DWs) in ferromagnetic nanowires. 
We demonstrate that the transverse component of nonlocal damping slows down the 
field-driven DW propagation and increases the Walker breakdown field whereas it is neglected in many previous works in literature. The experimentally measured DW 
mobility variation with the damping tuned by doping with heavy rare-earth elements that 
had discrepancy from micromagnetic simulation are now well understood with the 
nonlocal damping. Our results suggest that the nonlocal damping should be properly included as a prerequisite for quantitative studies of current-induced torques in noncollinear magnetization.
\end{abstract}

\pacs{75.78.-n, 75.60.Ch, 76.50.+g, 85.75.-d}
\maketitle

\section{Introduction}
Gilbert damping \cite{Gilbert}, which is spatially local and was
introduced to describe the energy dissipation in magnetization
dynamics, is a phenomenological parameter of magnetic materials
although a microscopic theory based on the spin-orbit interaction
and disorder scattering is available \cite{Kambersky1976}.
The energy dissipation plays an important role in both
current-driven \cite{Stiles2006} and field-driven \cite{Wang2009}
magnetization dynamics. For example, in the field-driven domain
wall (DW) propagation in a magnetic wire, the propagation speed
is proportional to the energy dissipation rate \cite{Wang2009}.
Thus a comprehensive understanding of dissipation (damping) in
magnetic materials is not only fundamentally interesting, but
also technologically important since the performance of many
spintronic devices such as race-track memory
\cite{Parkin2008} is directly related to the DW propagation speed.
In the past several decades, the progress in our understanding of the
damping has greatly advanced both theoretically and experimentally.
Theoretically, the Gilbert damping of real materials can be calculated
by using the torque-correlation model \cite{Gilmore2007,Kambersky2007,
Sakuma2012}, Kubo formalism \cite{Garate2009,Ebert2011,Mankovsky2013}
and the scattering approach \cite{Brataas2008,Starikov2010,YLiu2011} in
combination with the first-principles electronic structures.
In experiments, the value of Gilbert damping can be measured by
ferromagnetic resonance (FMR)
\cite{Kittel1948,Kobayashi2009,Langner2009,Liu2011,Weindler2014}.
For a nonuniform magnetic structure (noncollinear magnetization)
such as a DW in a ferromagnetic wire, FMR is not applicable and the
Gilbert damping is usually extracted via measuring the field-driven
DW velocity \cite{Ono1999, Atkinson2003,Wang2009, Moore2010}.
This technique is based on the following general features of the
field-driven DW propagation: below a critical field, a
DW propagates like a rigid body and its velocity is proportional to
the external field and inversely proportional to the Gilbert damping
\cite{Walker1974}. Surprisingly, the extracted Gilbert damping coefficient
of permalloy ($\mathrm{Ni}_{80}\mathrm{Fe}_{20}$) from field-driven DW
motion is three times larger than the value of the same material from FMR
measurement \cite{Weindler2014}. The enhanced damping in magnetic DW
has been attributed to the surface roughness of a ferromagnetic nanowire
in combination with texture-enhanced Gilbert damping arising from spin
pumping \cite{Weindler2014}.

Spin pumping was first proposed to understand the Gilbert damping
enhancement in a thin ferromagnetic film in contact with a nonmagnetic
metal \cite{Murakami2001}. A precessional magnetization $\mathbf m(t)$ pumps
an electron spin current of polarization $\mathbf j_{\rm pump}^s\sim
\mathbf m\times\partial\mathbf m/\partial t$ into the nonmagnetic metal
that dissipates via spin-flip scattering \cite{Tserkovnyak2002,Liu2014}.
Spin pumping does not have any observable effect in a homogeneous
magnetic structure because the net inflow/outflow spin current anywhere
in the system is zero due to a precise cancellation of the pumped
spin currents in opposite directions. In noncollinear magnetization,
the partial cancellation of the spatially dependent spin pumping
gives rise to a nonzero net inflow/outflow $-\nabla[\mathbf m
(\mathbf r) \times \partial \mathbf m(\mathbf r)/\partial t]$.
This results in an extra torque that has nonlocal damping in nature,
different from the local Gilbert damping due to spin-orbit interaction.
The total damping of a nonuniform magnetic structure such as a spin
spiral and a DW is enhanced through spin pumping
\cite{Foros2008,Tserkovnyak2009a,Zhang2009,Hankiewicz2008,Tserkovnyak2009b}
or spin wave emission \cite{Xiansi2012}.
The enhanced damping in nonuniform magnetic structure of permalloy has
been quantitatively calculated from the first-principles, which depends
not only on the magnetization gradient, but also on the particular
dynamical modes \cite{Yuan2014}.

The torque due to the pumped spin current from precessional noncollinear magnetization
can be decomposed into a longitudinal
and a transverse components, which were independently predicted by
several groups around the same time. Foros {\it et al.} \cite{Foros2008} and
Zhang {\it et al.} \cite{Zhang2009} showed that the longitudinal
spin current increases the effective damping in spin spirals and DWs.
Hankiewicz {\it et al.} \cite{Hankiewicz2008,Tserkovnyak2009b} discovered that
the transverse spin current influences the dissipation of spin waves. Many previous 
works in literature only include the longitudinal component. \cite{Weindler2014,Kim2011,Moon2012,Kim2012}
For the field-driven transverse DW motion, the first-principles calculation
showed that the transverse and longitudinal spin currents are
responsible for the rigid-body motion below a critical field and the
oscillatory motion above the critical field, respectively \cite{Yuan2014}.
For more complicate magnetic structures with noncollinear magnetization,
e.g. vortex DWs, how this extra nonlocal damping influences magnetization
dynamics is not entirely clear. For example, it was predicted that the
longitudinal component of the pumped spin current has no effect on
steady-state DW motion \cite{Zhang2009}.
On the other hand, micromagnetic simulation showed that the longitudinal
component can slow down the field-driven DW propagation as what
was observed in experiments \cite{Weindler2014}. The inconsistency in the
theory and experiment requires a better understanding of the effects of
the nonlocal damping on the DW motion.

In this paper, the complete form of the nonlocal damping is derived from physically transparent spin pumping formalism and is incorporated into a generalized Thiele equation (GTE), which describes the steady-state motion of a nonuniform magnetic structure. We focus on the
consequences on propagation of different types of DWs resulting from the nonlocal damping.
The solutions of the GTE explicitly show the role played by the nonlocal damping in DW motion, where the longitudinal and transverse components are equally important in general. For the steady motion of a DW driven by an external field, the longitudinal component vanishes while the transverse one slows down the DW velocity and increases the Walker breakdown field. The experimentally observed dependence of the DW
mobility on the damping parameter, which were systematically lower than the values from micromagnetic simulations, can be understood by including the nonlocal damping. As a prerequisite, quantitative extraction of current-induced torques in noncollinear magnetization dynamics relies on an accurate description of nonlocal damping in analytical models and micromagnetic simulations. The paper is organized as follows. In Sec.~\ref{sec:II}, we derive the
nonlocal damping and incorporate it into the GTE. The GTE for both
transverse and vortex DWs in a magnetic nanowire is then solved in Sec.~
\ref{sec:III} under a static external magnetic field along the wire.
The solution is also used to determine the strength of the nonlocal
damping from available experimental data. The conclusions are summarized
in Sec.~\ref{sec:IV}. In Appendix~\ref{app:mixing}, we provide a detailed
derivation of the spin pumping and the resulting damping torque of
noncollinear magnetization in a ferromagnetic nanowire.

\section{General formalism\label{sec:II}}
\subsection{Nonlocal damping torque}
We consider a nonuniform magnetic structure in a nanowire as
schematically shown in Fig. 1. The $x-$, $y-$, and $z-$axes are
along the length, width and thickness directions, respectively.
A spin current polarized along $\sim\mathbf m(\mathbf r,t)\times\partial_t\mathbf
m (\mathbf r,t)$ is pumped out towards all directions \cite{Tserkovnyak2002}
as magnetization $\mathbf m(\mathbf r, t)$ changes with time.
While the detailed derivation is given in Appendix~\ref{app:mixing}, the resulting spin current in a noncollinear magnetization reads
\begin{eqnarray}
\mathbf j^s_i(\mathbf r,t)&=&-\frac{\hbar}{4\pi}\Gamma_{\uparrow\downarrow}
\partial_i\left[\mathbf m (\mathbf r,t)\times\partial_t\mathbf m (\mathbf r,t)\right]\nonumber\\
&=&-\frac{\hbar}{4\pi}\Gamma_{\uparrow\downarrow}\partial_i
\mathbf m (\mathbf r,t)\times\partial_t\mathbf m (\mathbf r,t)\nonumber\\
&&-\frac{\hbar}{4\pi}\Gamma_{\uparrow\downarrow}\mathbf m(\mathbf r,t)
\times\partial_i\partial_t\mathbf m (\mathbf r,t)\nonumber\\
&\equiv&\mathbf j^{s\parallel}_i(\mathbf r,t)+\mathbf j^{s\perp}_i(\mathbf r,t).
\label{eq:is}
\end{eqnarray}
Here $i=x,y,z$ denotes the propagation direction of the spin current.
$\Gamma_{\uparrow\downarrow}$ is the (intralayer) spin-mixing
conductivity that has the dimension of the inverse of length.
Its relation to the conventional spin-mixing conductance is disucssed in
Appendix~\ref{app:mixing}. The longitudinal spin current (LSC)
$\mathbf j^{s\parallel}_i=-\frac{\hbar}{4\pi}\Gamma_{\uparrow\downarrow}
\partial_i\mathbf m \times\partial_t\mathbf m$ is always aligned with the
local magnetization $\mathbf m$ since both $\partial_i\mathbf m$
and $\partial_t \mathbf m$ are perpendicular to $\mathbf m$.
$\mathbf j^{s\perp}_i=-\frac{\hbar}{4\pi}\Gamma_{\uparrow\downarrow}
\mathbf m \times\partial_i\partial_t\mathbf m $ is the transverse spin
current (TSC), which is perpendicular to the local magnetization $\mathbf m$.

In early works \cite{Foros2008,Zhang2009} with the adiabatic approximation,
in which the polarization of the spin current is assumed to align with
local magnetization, the transverse spin current is artificially neglected.
This approximation is also used in some later works \cite{Weindler2014,Kim2011,
Moon2012,Kim2012} without a proper justification. Taking permalloy as an example,
the first-principles calculations showed that the spin diffusion length
and the transverse spin coherent length are 5.5~nm \cite{Starikov2010}
and 13.1~nm \cite{Yuan2014}, respectively, at low temperature.
These lengths are not much smaller than the width of DWs
\cite{McMichael1997,Nakatani2005}, as required for the adiabatic approximation.
Thus there are no reasons that the TSC in real materials can be neglected.
Indeed, we will see that the TSC substantially influences the magnetization
dynamics in the nonuniform magnetic structures, especially for the
steady-state DW motion below the critical field.

\begin{figure}[t]
\centering
\includegraphics[width=\columnwidth]{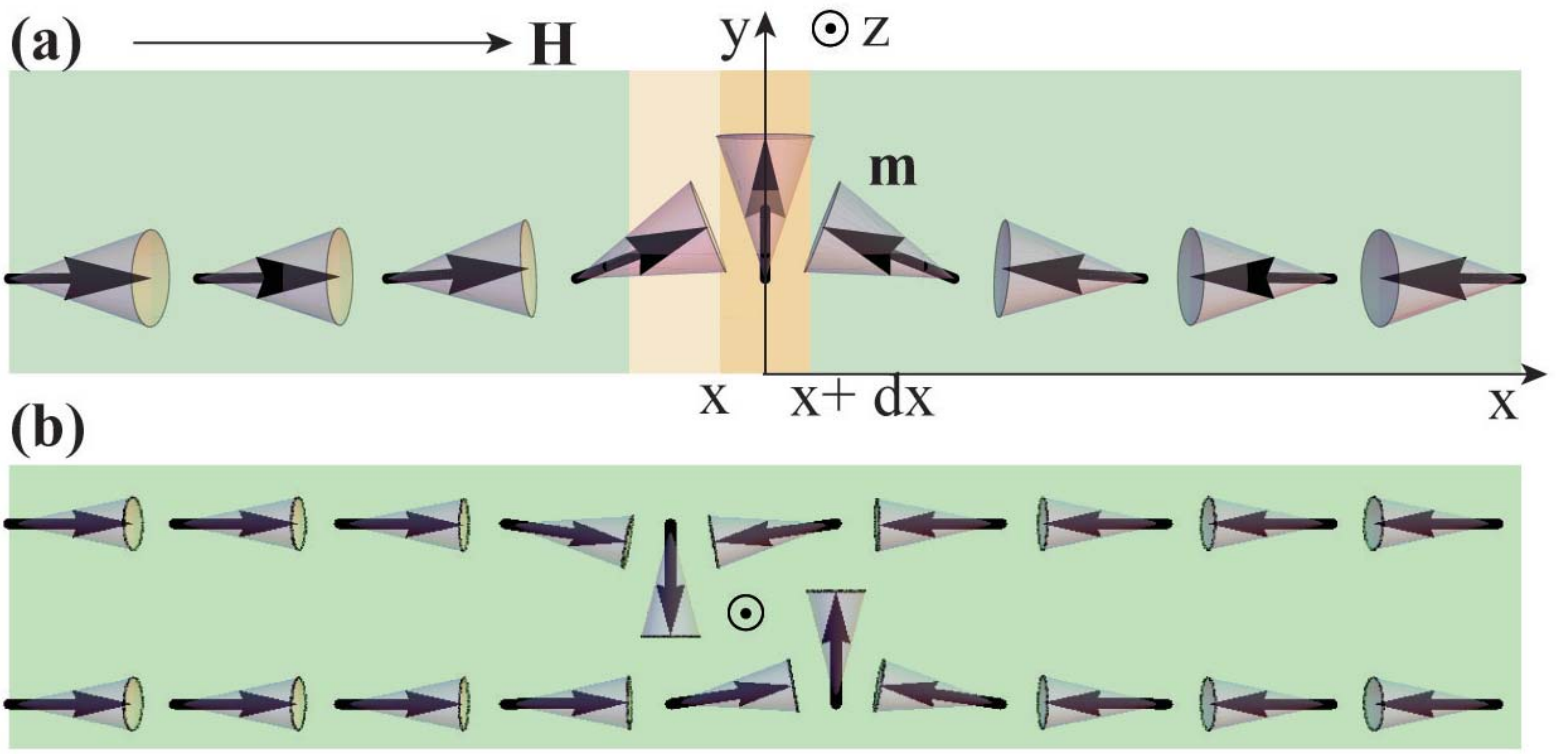}
\caption{(a) Top view of a head-to-head transverse DW in
a magnetic nanowire. The thick arrows denote the directions of
magnetization $\mathbf{m}$. The $x-$,$y-$,$z-$axes are
along the length, width and thickness directions of the nanowire, respectively.
(b) Top view of a vortex DW in a nanowire. The circle in the center denotes the vortex core.}
\label{fig1}
\end{figure}

The dissipative torque generated by the precessional motion of noncollinear
magnetization is given by the divergence of the spin current in Eq.~(\ref{eq:is}),
\begin{eqnarray}
\bm{\tau}_{\rm damping} = \frac{\gamma}{M_s} \partial_i
\left(\mathbf j^{s\parallel}_i+\mathbf j^{s\perp}_i\right)
=\bm{\tau}_{\rm LSC}+\bm{\tau}_{\rm TSC},\label{eq:damping}
\end{eqnarray}
where $\gamma=g\mu_B/\hbar$ is the gyromagnetic ratio in terms of
Land{\'e} $g$ factor and Bohr magneton $\mu_B$ and $M_s$ is the
saturation magnetization. Einstein summation is assumed in this paper
unless it is stated otherwise. The total torque can also be decomposed into
the longitudinal and transverse components, $\bm{\tau}_{\rm LSC}$ and
$\bm{\tau}_{\rm TSC}$. Specifically, we have
\begin{eqnarray}
\bm{\tau}_{\mathrm{TSC}} &=&- \frac{g\mu_B\Gamma_{\uparrow\downarrow}}
{4\pi M_s} \partial_i ( \mathbf{m} \times \partial_i \partial_t \mathbf{m} )
\nonumber\\
&=&\frac{g\mu_B\Gamma_{\uparrow\downarrow}}{4\pi M_s} \left[(\mathbf{m}
\cdot \partial_i  \partial_t \mathbf{m}) {\mathbf{m}}
\times \partial_i  \mathbf{m}
- \mathbf{m} \times
\partial^2_i \partial_t \mathbf{m}\right],\nonumber\\
\label{eq:tsc}
\end{eqnarray}
and
\begin{eqnarray}
\bm{\tau}_{\mathrm{LSC}} &=&-\frac{g\mu_B\Gamma_{\uparrow\downarrow}}
{4\pi M_s} \partial_i \left(\partial_i  \mathbf{m} \times
\partial_t \mathbf{m} \right) \nonumber\\
&=&\frac{g\mu_B\Gamma_{\uparrow\downarrow}}{4\pi M_s}\mathbf{m} \times
(\mathbf{A} \cdot \partial_t \mathbf{m} ),\label{eq:lsc}
\end{eqnarray}
where $\mathbf{A}$ is dyadic tensor
\begin{equation}
\mathbf{A}= (\mathbf{m} \times \partial_i \mathbf{m} )(\mathbf{m}
\times \partial_i \mathbf{m}).
\end{equation}
Equation~(\ref{eq:lsc}) reproduces the tensor form obtained in
Ref.~\onlinecite{Zhang2009}. Alternatively, the second term of
$\bm{\tau}_{\rm TSC}$ could also be obtained through an expansion of
the Gilbert damping term to the second-order spatial derivative of
magnetization \cite{Tserkovnyak2009b} or through the phenomenological
Landau-Lifshitz-Baryakhtar equation \cite{Wang2015}.
In a very weak texture like a spin wave, where the precessing
magnetization deviates slightly from the equilibrium direction with a
cone angle $\theta$, the energy dissipation due to the first term of
$\bm{\tau}_{\rm TSC}$ is proportional to $\sin^4\theta$ while the second
term is proportional to $\sin^2\theta$ \cite{Yuan2014}.
This is the reason why the first term in Eq.~(\ref{eq:tsc}) can be
neglected in spin wave dynamics \cite{Hankiewicz2008,Tserkovnyak2009b}.
It should not be neglected in strong magnetization textures like
a transverse DW \cite{Yuan2014}.

\subsection{Generalized Thiele equation}
To describe the dynamics of noncollinear magnetization, we add the
two torques $\bm{\tau}_\mathrm{LSC}$ and $\bm{\tau}_\mathrm{TSC}$
into the Landau-Lifshitz-Gilbert (LLG) equation, i.e.
\begin{equation}
\partial_t \mathbf{m} =-\gamma\mathbf{m} \times \mathbf{H}_{\rm eff}
+\alpha \mathbf m \times \partial_t \mathbf{m}  + \bm{\tau}_\mathrm{LSC}
+ \bm{\tau}_\mathrm{TSC}.
\label{llgt}
\end{equation}
Here $\mathbf{H}_{\rm eff}$ is the effective field that the external applied field, the exchange field, anisotropy and demagnetization fields. $\alpha$ is the usual local
Gilbert damping coefficient. For simplicity, we rewrite  the LLG
equation in a more compact form as \cite{Note0}
\begin{equation}
\partial_t \mathbf{m} =-\mathbf{m} \times \left ( \gamma\mathbf{H}_{\rm eff}
-\alpha \partial_t \mathbf{m} \right ) - \eta\nabla^2(\mathbf{m}
\times \partial_t \mathbf{m} ),
\label{llg}
\end{equation}
where we have defined $\eta\equiv g\mu_B\Gamma_{\uparrow\downarrow}/(4\pi M_s)$
representing the strength of nonlocal damping.
Following the Thiele \cite{Thiele1973} analysis for the field-driven
rigid DW motion and by $\mathbf m\times\left[\mathrm{Eq.~(\ref{llg})}
\right]\cdot \partial_i \mathbf{m}$, we have
\begin{eqnarray}
(\mathbf{m} \times \partial_t \mathbf{m}) \cdot \partial_i \mathbf{m}
&=&\gamma\mathbf{H}_{\rm eff} \cdot {\partial_i \mathbf{m}}-
\alpha (\partial_t \mathbf{m}) \cdot(\partial_i \mathbf{m}) \nonumber \\
&&+ \eta(\partial_i \mathbf{m} \times \mathbf{m})
\cdot \nabla^2(\mathbf{m} \times \partial_t \mathbf{m}).
\label{llg2}
\end{eqnarray}

For the rigid DW motion, the magnetization is only the function of
$\mathbf r-\mathbf vt$ \cite{Thiele1973, Slonczewski1979} , i.e.
\begin{equation}
\theta (\mathbf r) =\theta (\mathbf r-\mathbf vt),~\varphi(\mathbf r)
= \varphi(\mathbf r-\mathbf vt),
\end{equation}
where $\theta$ and $\varphi$ are, respectively, the polar and azimuthal
angles of
$\mathbf{m}=(\sin\theta\cos\varphi,\sin\theta\sin\varphi,\cos\theta)$ in
spherical coordinates and $\mathbf{v}$ is velocity of the magnetic structure.
Then the spatial and time derivatives of magnetization can be written as
\begin{eqnarray}
\partial_i\mathbf m&=&\partial_i\theta\hat{\theta}+\sin\theta
\partial_i\varphi\hat{\varphi},\nonumber\\
\partial_t\mathbf m&=&\partial_t\theta\hat{\theta}+\sin\theta
\partial_t\varphi\hat{\varphi}\nonumber\\
&=&(-\mathbf v\cdot\nabla\theta)\hat{\theta}
+\sin\theta(-\mathbf v\cdot\nabla\varphi)\hat{\varphi}.
\label{steady}
\end{eqnarray}

Substituting Eq.~(\ref{steady}) into Eq.~(\ref{llg2}), we obtain
\begin{widetext}
\begin{eqnarray}
\sin \theta \left[\left (\mathbf{v} \cdot \nabla \varphi \right)\partial_i \theta
- \left(\mathbf{v} \cdot \nabla \theta \right)\partial_i\varphi \right]
&=& \gamma\partial_i(\mathbf{m} \cdot \mathbf{H}_{\rm ext})
 + \alpha \mathbf v\cdot\left ( \nabla \theta \partial_i \theta
 + \sin^2 \theta \nabla \varphi \partial_i \varphi \right )\nonumber\\
&&- \eta \left[ \left(-\partial_i\theta\hat{\varphi}+\sin\theta\partial_i\varphi\hat{\theta}\right)
 \cdot \nabla^2\left(\sin\theta\nabla\varphi\hat{\theta}-\nabla\theta\hat{\varphi}\right)
 \right] \cdot \mathbf v.\label{eq:thielefull}
\end{eqnarray}
\end{widetext}
Here $\mathbf{H}_{\rm ext}$ is the external magnetic field.
Multiplying $M_s$ to the both sides of Eq.~(\ref{eq:thielefull}) and
integrating over the whole nanowire, the generalized Thiele equation becomes
\begin{equation}
\mathbf{F} +\mathbf{G}\times \mathbf{v}+ \alpha\mathbf D
 \cdot \mathbf{v} + {\eta }\mathbf{D'} \cdot \mathbf{v}= 0,
\label{thiele}
\end{equation}
where the vectors $\mathbf F$ and $\mathbf G$ and the tensors
$\mathbf D$ and $\mathbf D'$ are, respectively,
\begin{eqnarray}
\mathbf{F} &=& M_s \int \nabla \left(-\mathbf{m}\cdot
\mathbf{H}_{\rm ext}\right)d^3r, \nonumber\\
\mathbf{G} &=& -\frac{M_s}{\gamma} \int \left(\sin \theta \nabla \theta
\times \nabla \varphi\right) d^3r, \nonumber\\
\mathbf{D} &=& -\frac{M_s}{\gamma} \int \left(\nabla \theta \nabla \theta
+ \sin^2 \theta \nabla \varphi \nabla \varphi \right) d^3r, \nonumber\\
\mathbf{D'} &=&\frac{M_s}{\gamma} \int \left(\sin\theta\nabla\varphi
\hat{\theta}-\nabla \theta\hat{\varphi}\right)
\cdot \nabla^2\left(\sin\theta\nabla\varphi\hat{\theta}
-\nabla\theta\hat{\varphi}\right) d^3r.\nonumber\\
 \label{eq:fgdd}
\end{eqnarray}
The nonlocal damping appears in the new dissipation term
$\eta \mathbf{D}' \cdot \mathbf v$ in Eq.~(\ref{thiele}).
The original Thiele equation \cite{Thiele1973} is reproduced for $\eta = 0$.

\section{Field-driven DW motions\label{sec:III}}
To explicitly see the effects of nonlocal damping on the DW motion,
we apply the GTE (\ref{thiele}) to the propagation of transverse and vortex DWs.
The analytical results for transverse DWs are compared with micromagnetic
simulations of the LLG Eq.~(\ref{llgt}). We will also use the available
experimental data in literature to extract the nonlocal damping coefficient
$\eta$.

\subsection{Transverse DWs}
A transverse DW is energetically preferred in relatively narrow
and thin nanowires \cite{McMichael1997, Nakatani2005}.
We consider a head-to-head N{\'e}el DW in a ferromagnetic nanowire
of thickness $T$ and width $W$, as illustrated in Fig.~\ref{fig1}(a).
The magnetization  $\mathbf{m}(\mathbf r-\mathbf vt)$ is only a function of $x-v_xt$,
and $\mathbf m(-\infty)=-\mathbf m(+\infty)=\hat{\mathbf x}$.
An external field $\mathbf{H}_{\rm ext}= H\hat{\mathbf x}$ is
applied that drives the DW to propagate along $+x$ direction.
Under these conditions, Eq.~(\ref{eq:fgdd}) gives,
\begin{eqnarray}
\mathbf F&=&-TW M_s H\hat{\mathbf x}\int dx\,\partial_x m_x(x)
=2TW M_s H\hat{\mathbf x},\nonumber\\
\mathbf G&=&0,\nonumber\\
\mathbf D&=&-\frac{TWM_s}{\gamma} \hat{\mathbf x}\hat{\mathbf x}\int dx\,\left[\left(\partial_x\theta\right)^2+\sin^2\theta
\left(\partial_x\varphi\right)^2\right]\nonumber\\
&=&-\frac{TWM_s}{\gamma} \hat{\mathbf x}\hat{\mathbf x}
\int dx\,\left[\partial_x\mathbf m(x)\right]^2,\nonumber\\
\mathbf D'&=&\frac{TWM_s}{\gamma}\hat{\mathbf x}\hat{\mathbf x}
\int dx\,d'_{xx}[\theta(x),\varphi(x)],\label{eq:tw}
\end{eqnarray}
where $d'_{xx}$ is defined as
\begin{equation}
d'_{xx}=\left(\sin\theta\partial_x\varphi\hat{\theta}
-\partial_x\theta\hat{\varphi}\right)
 \cdot \partial^2_x\left(\sin\theta\partial_x\varphi\hat{\theta}
 -\partial_x\theta\hat{\varphi}\right).
 \label{eq:dp}
\end{equation}
Substituting Eq.~(\ref{eq:tw}) into the GTE (\ref{thiele}),
we find the DW steady-state velocity
\begin{equation}
v_x = \frac{2\gamma H}{\int dx\left[\alpha (\partial_x\mathbf m)^2
-\eta d'_{xx}\right]}.\label{eq:vx1}
\end{equation}

For a N{\'e}el DW centered at $x_0$ and with a width $\lambda$,
the polar and azimuthal angles of the magnetization are given by
the Walker profile \cite{Walker1974},
\begin{eqnarray}
&&\theta(x)=\frac{\pi}{2},\nonumber\\
&&\varphi(x)=\pi-\arccos\left[\tanh\left(\frac{x-x_0}{\lambda}\right)\right].
\end{eqnarray}
The field-driven DW velocity of Eq.~(\ref{eq:vx1}) can be explicitly calculated,
\begin{equation}
v_x = \frac{\gamma H\lambda}{\alpha + \eta /(3\lambda^2)}.
\label{eq:vxtw}
\end{equation}
The effective damping of the Walker DW with the nonlocal damping is
$\alpha_{\rm eff}=\alpha+\eta /(3\lambda^2)$ in agreement with the
calculated in-plane damping of transverse DWs using the first-principles
scattering theory \cite{Yuan2014}.
The DW propagation is slowed down by the nonlocal damping for
a given field.

The so-called Walker breakdown $H_{\rm W}$ \cite{Walker1974},
above which the solution for a rigid DW motion does not exist, is
$H_{\rm W}=\alpha K_z/(\mu_0M_s)$ \cite{Walker1974,Wang2009}  in
a one-dimensional model and in the absence of the nonlocal damping,
where $K_z$ is the total magnetic anisotropy energy along the hard
axis including both the magnetocrystalline and shape anisotropy.
In the presence of the nonlocal damping, the effective Gilbert damping
of a Walker DW becomes $\alpha+\eta/(3 \lambda^2)$ and the Walker
breakdown field becomes larger,
\begin{equation}
H_{\rm W}=\frac{K_z(3\alpha+\eta/\lambda^2)}{3\mu_0 M_s}.\label{eq:hw}
\end{equation}
The corresponding velocity at the breakdown field can be calculated
from Eq.~(\ref{eq:vxtw}) and Eq.~(\ref{eq:hw}),
\begin{equation}
v_x^{\rm W}=\frac{\gamma K_z \lambda}{\mu_0M_s}, \label{eq:vxmaxtw}
\end{equation}
which is not affected by the nonlocal damping.

\begin{figure}[t]
\centering
\includegraphics[width=\columnwidth]{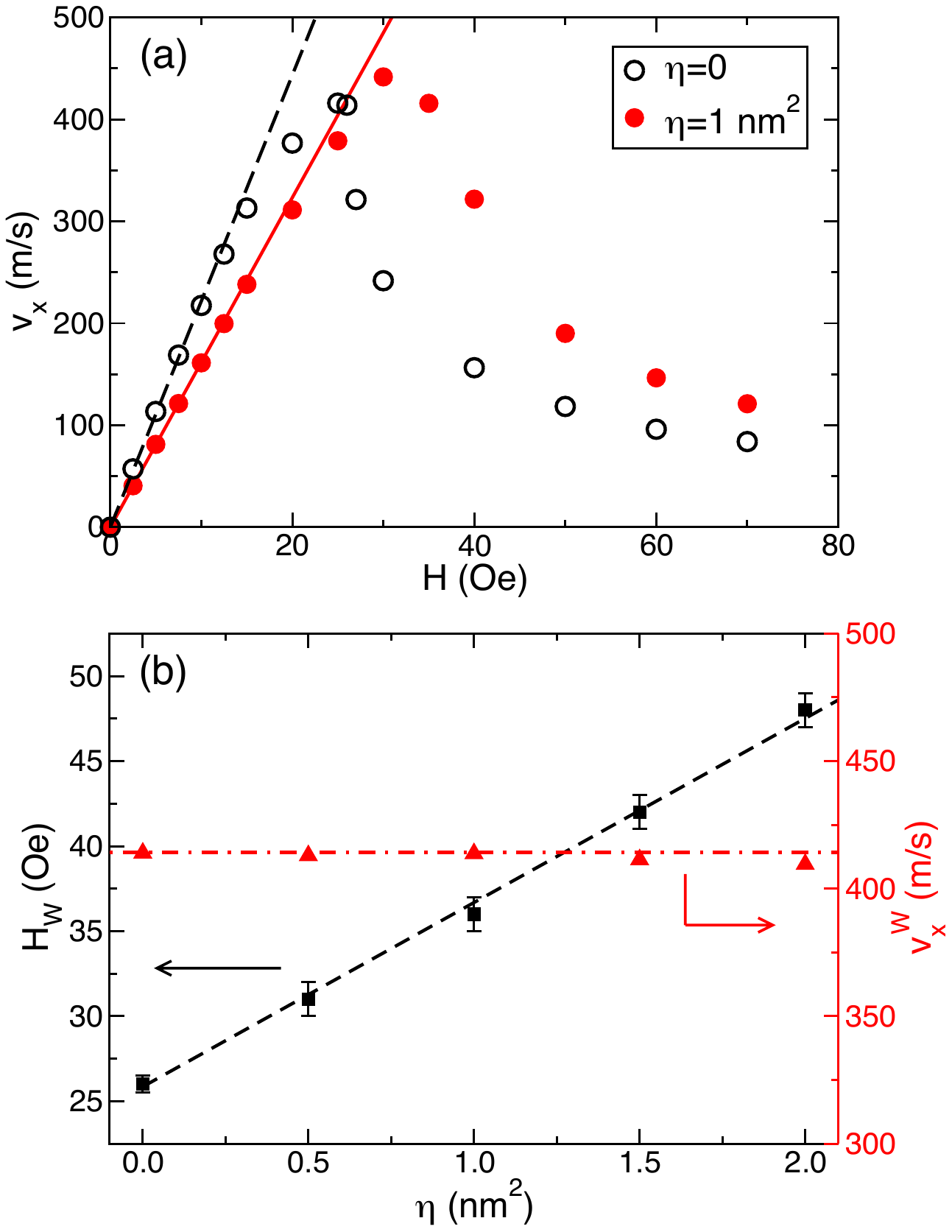}
\caption{(a) The velocity of a transverse DW as a function of the external
field from numerical simulations without the non-local damping (black open
circles for $\eta=0$) and with $\eta =1.0\ \mathrm{nm^2}$ (red solid circles).
The black dashed and red solid lines illustrate Eq.~(\ref{eq:vxtw}) with $\eta=0$
and $\eta =1.0\ \mathrm{nm^2}$, respectively. (b) The Walker breakdown field
$H_{\rm W}$ (left axis) and the velocity of DW motion $v_x^{\rm W}$ at the
Walker breakdown field as a function of the nonlocal damping strength $\eta$.
The black dashed line is a linear least squares fitting to the black solid
squares, which gives the total anisotropy
$K_z=(2.1\pm0.1)\times10^5~\mathrm J/\mathrm m^3$.
The red dashed-dotted line is the calculated velocity
at the breakdown field using Eq.~(\ref{eq:vxmaxtw}) with the fitted $K_z$.}
\label{fig2}
\end{figure}
The LLG equation (\ref{llgt}) is numerically solved for a transverse
DW in a nanowire of 4 nm thick and 16 nm wide under an external field.
The system is discretized into uniform meshes of size 4~nm.
The permalloy parameters are used with the saturation magnetization
$M_s = 8 \times 10^5~\mathrm{A/m}$, the exchange stiffness $A=1.3 \times
10^{-11}~\mathrm{J/m}$, the in-plane crystalline anisotropy (along
$x$) $K_c=500~\mathrm{J/m}^3$ and Gilbert damping $\alpha=0.01$.
The shape anisotropy is implicitly taken into account via including
the dipole-dipole interaction in the simulation.
Using the above material parameters, we obtain the static DW width
$\lambda\approx12.6$~nm. When an external field is applied, the width decreases
\cite{Walker1974} and approaches 9 nm near the breakdown field $H_{\rm W}$.

Figure~\ref{fig2}(a) shows the field-dependence of DW propagation velocity
$v_x$ along the $+x$ direction. At a low external field, $v_x$ is
proportional to $H$, following Eq.~(\ref{eq:vxtw}) (straight lines).
For the external field larger than 20~Oe, the numerical velocity
is slightly lower than the analytical value.
This is because the Walker solution becomes unstable as the field is close
to the breakdown field and spin wave emission in this regime would
further slow down the DW propagation \cite{Xiansi2012, BHu2013}.

The critical field $H_{\rm W}$ that is obtained from numerical
simulations is plotted in Fig.~\ref{fig2}(b) as a function of $\eta$.
The linear relation in Eq.~(\ref{eq:hw}) is reproduced
and a linear least squares fitting yields the intercept
$H_{\rm W}(\eta=0)=25.9\pm0.3~$Oe and the slope
$H_{\rm W}/\eta=10.8\pm0.3~\mathrm{Oe}/\mathrm{nm}^2$.
Using Eq.~(\ref{eq:hw}), the values of both the intercept and slope
consistently lead to the effective anisotropy energy
$K_z=(2.1\pm0.1)\times 10^5~\mathrm{J}/\mathrm m^3$.
Then we are able to calculate the analytical value of the DW velocity at the breakdown
field using Eq.~(\ref{eq:vxmaxtw}), $v_x^{\rm W}=414~$m/s,
as shown by the red  horizontal dashed-dotted line in Fig.~\ref{fig2}(b).
The values of $v_x^{\rm W}$ obtained from the numerical simulation plotted in Fig.~\ref{fig2}(b) are
in very good agreement with the analytical value.

\subsection{Vortex DWs\label{sec:vdw}}
A vortex DW is energetically more stable in relatively wide and thick
nanowires \cite{McMichael1997, Nakatani2005}. Here a vortex DW is modeled by
an inner vortex core with out-of-plane magnetization and an outer curling
structure \cite{yuan2016}. The spatial dependence of the polar and azimuthal
angles can be analytically described as \cite{Huber1982, He2006}
\begin{eqnarray}
&&\theta(x,y) = \left\{
\begin{array}{ll}
  2\arctan\left(\displaystyle\frac{r}{r_c}\right),  & r\leq r_c,\\
  \displaystyle\frac{\pi}{2}, & r>r_c,
\end{array} \right. \nonumber\\
&&\varphi(x,y) = \arg[(x-x_0)+i(y-y_0)]+\frac{\pi}{2},
\end{eqnarray}
where $(x_0,y_0)$ is the center of the vortex core and
$r=\sqrt{(x-x_0)^2+(y-y_0)^2}$.
$r_c$ is the radius of vortex core that is comparable to the exchange
length ($\sim 5$ nm for permalloy).

The motion of a vortex DW under an external field $\mathbf H_{\rm ext}=
H\hat{\mathbf x}$ can be described by applying the GTE (\ref{thiele}).
Because $r_c$ is much smaller than the outer curling structure (several tens
to hundreds of nanometers) of a vortex DW, the dominant contribution to
$\mathbf D$ and $\mathbf D'$ comes from the spins in the curling structure,
\begin{eqnarray}
\mathbf{D} &=& - \frac{M_s}{\gamma}\int (\sin^2 \theta \nabla \varphi
\nabla \varphi )d^3r \nonumber\\
&=& -\frac{TM_s}{\gamma} \int \left (\frac{\partial \mathbf{m}}{\partial x}\right )^2dxdy
\left(\mathbf{\hat{x}\hat{x}} + \mathbf{\hat{y}\hat{y}}\right)\nonumber\\
&=& - \frac{\pi TM_s}{ \gamma}\ln\frac{W}{2r_c} \left(\mathbf{\hat{x}\hat{x}}
+ \mathbf{\hat{y}\hat{y}}\right).
\end{eqnarray}
Here we have used $W/2$ as the outer radius of the vortex DW.
The tensor $\mathbf{D}'$ is difficult to derive analytically
due to the high order spatial derivatives of $\theta$ and
$\varphi$ but it is well converged numerically,
\begin{equation}
\mathbf D'=-\frac{3.3TM_s}{\gamma r_c^2}\left(\mathbf{\hat{x}\hat{x}}
+ \mathbf{\hat{y}\hat{y}}\right).
\end{equation}

Similarly, it is a reasonable approximation to neglect the contribution
of the vortex core to force $\mathbf F$. So we have
\begin{eqnarray}
\mathbf{F}&=& - TM_sH\int \nabla ( \sin \theta \cos \varphi)d^2r \nonumber\\
       &\approx& TM_sH\int_0^{2\pi} \int_{r_c}^{R} \left(\frac{\sin^2 \phi}{r},
       \frac{-\cos \phi \sin \phi}{r}\right)rdr d\phi \nonumber\\
       &=& \frac{\pi  WTM_s H}{2}\mathbf{\hat{x}}.
\end{eqnarray}
Following He {\it et al.}, \cite{He2006} we add a restoring force
that is linear in the transverse displacement $\delta y$ of the
vortex core from the nanowire center,
$\mathbf F^{\rm re}=-\hat{\mathbf{y}}\kappa \delta y$, where the
coefficient $\kappa$ depends on the magnetic anisotropy and the
equilibrium width of the vortex DW. \cite{He2006}

For the gyrovector $\mathbf G$, the spins outside the vortex core
do not contribute because of $\theta =\pi/2$. The integral
over the vortex core can be analytically evaluated as \cite{He2006}
\begin{equation}
\mathbf{G} = -\frac{2 \pi  TM_s}{\gamma}  \mathbf{\hat{z}}.
\end{equation}

Substituting the expressions of $\mathbf{F}, \mathbf{G}, \mathbf{D}$
and $\mathbf{D'}$ into GTE~(\ref{thiele}), the coupled equations become
\begin{eqnarray}
\frac{\gamma H W}{2} +2 v_y-\left(\alpha \ln\frac{W}{2r_c} +
\frac{3.3\eta}{\pi r_c^2}\right) v_x &=&0,\nonumber\\
-\frac{\gamma\kappa \delta y}{\pi TM_s} - 2 v_x -
\left( \alpha \ln\frac{W}{2r_c} + \frac{3.3\eta}{\pi r_c^2}\right)v_y &=&0.
\end{eqnarray}
Under an external field, the vortex core moves  both along the field
direction (longitudinal) and transverse to the field ($y$ direction)
due to the gyro force. Then the vortex structure is deformed with
the core displacement $\delta y$ (from the nanowire centre).
At steady-state motion, the gyro effect is balanced by the restoring
force $\mathbf F^{\rm re}$, i.e. $v_y=0$. Then the vortex core only
moves in the longitudinal direction and the velocity can be obtained,
\begin{equation}
v_x=\frac{\gamma H\frac{W}{2}}{\alpha\ln\frac{W}{2r_c}+\frac{3.3\eta}
{\pi r_c^2}} \equiv\frac{\gamma H\lambda_{\rm eff}}{\alpha_{\rm eff}},
\label{eq:vxvw}
\end{equation}
where the effective width is defined as
\begin{equation}
\lambda_{\rm eff}\equiv\frac{W}{2\ln[W/(2r_c)]},
\end{equation}
and the effective damping is
\begin{equation}
\alpha_{\rm eff}\equiv\alpha+\frac{3.3\eta}{\pi r_c^2\ln[W/(2r_c)]}.
\label{eq:alphaeff}
\end{equation}
According to Eq.~(\ref{eq:alphaeff}), the nonlocal damping enhances the
effective damping of a vortex DW. The enhancement depends on the strength
of nonlocal damping ($\eta$), the size of vortex core ($r_c$)
and the width of nanowire ($W$).

Similar to the case of transverse DWs, the Walker breakdown field
increases in the presence of the nonlocal damping. For steady-state motion
with $v_y=0$, the transverse displacement of the vortex core is given by
\begin{equation}
\delta y =- \frac{2\pi T M_s H \lambda_{\rm eff}}{\alpha_{\mathrm{eff}}\kappa}.
\label{eq:dy}
\end{equation}
Equation~(\ref{eq:dy}) indicates that the transverse displacement
increases with the external magnetic field. When the vortex core
reaches the edge of the nanowire $\delta y=-W/2$, one has
the Walker breakdown field
\begin{equation}
H_\mathrm{W}=\frac{\alpha_{\mathrm{eff}}\kappa W}{4\pi T M_s \lambda_{\rm eff}}.
\end{equation}
With the nonlocal damping, the breakdown field increases and is
proportional to the effective damping $\alpha_{\rm eff}$.
The longitudinal velocity at the breakdown field is independent of
the nonlocal damping as in the case of transverse DWs.

\begin{figure}
\centering
\includegraphics[width=\columnwidth]{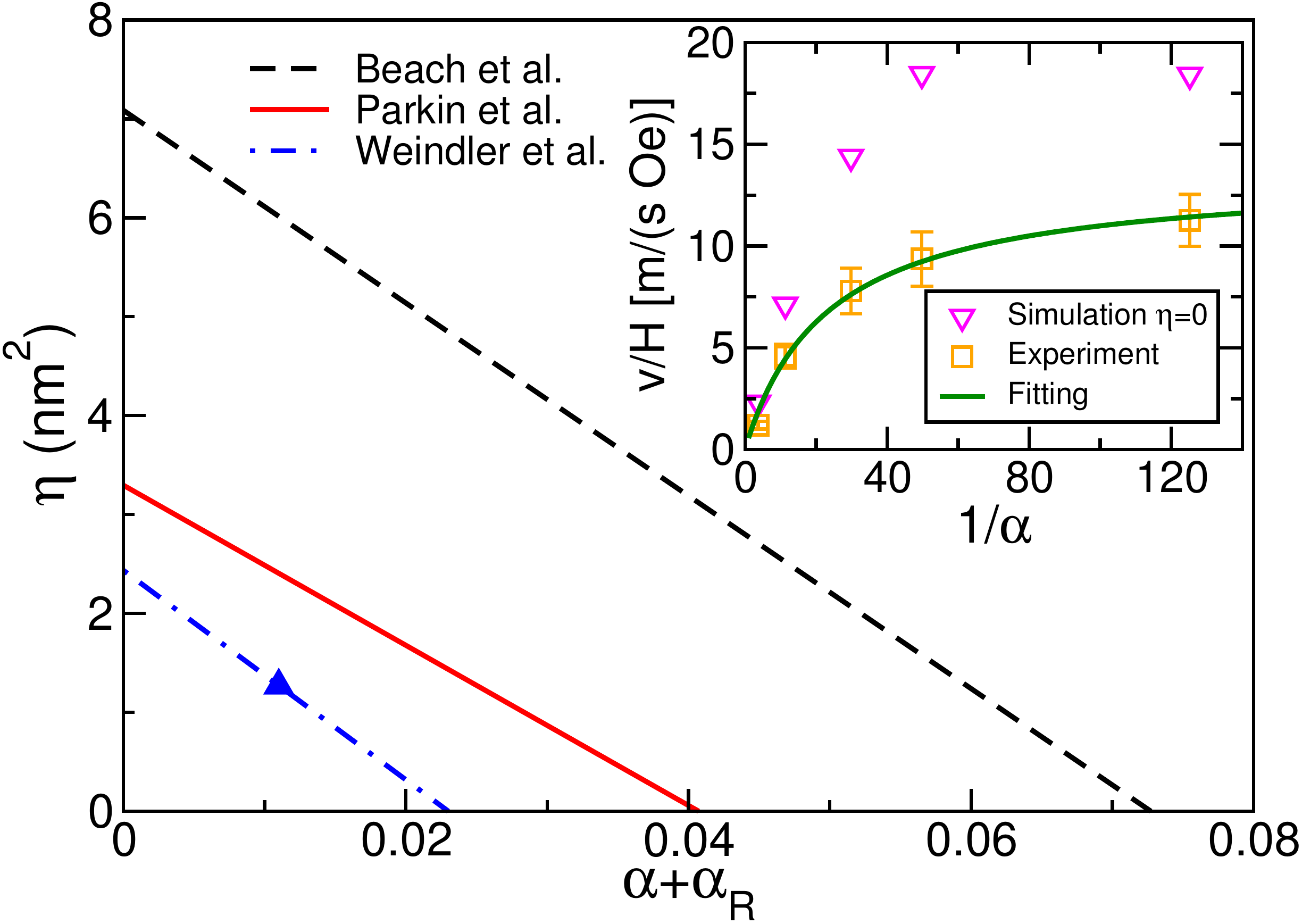}
\caption{Calculated $\eta$ as a function of Gilbert damping parameter
including the contributions from spin-orbit interaction $\alpha$ and
roughness $\alpha_R$, for the experimental DW velocities driven by an external
field taken from Ref.~\onlinecite{Beach2005} (black dashed line),
Ref.~\onlinecite{Parkin2008} (red solid line) and Ref.~\onlinecite
{Weindler2014} (blue dashed-dotted line). In Ref.~\onlinecite{Weindler2014},
$\alpha+\alpha_R=0.011$ is determined by fitting experimental data to
the simulations with various DW pinning strengths, and the corresponding
nonlocal damping strength $\eta=1.3$~nm$^2$ is obtained \cite{Note1} (blue solid triangle).
Inset: Gilbert damping dependence of DW mobility of Ho-doped permalloy
nanowires. The orange open squares are experimental data and the magenta
open triangles are the simulation without $\eta$ \cite{Moore2010}.
The green solid line shows a least squares fitting using Eq.~(\ref{eq:vxvw}),
which yields the upper bound of the nonlocal damping strength
$5.1\pm 1.6~\mathrm{nm}^2$.}
\label{fig3}
\end{figure}
Determining the value of $\eta$ from field-driven DW motion and Eq.~(\ref{eq:vxvw}) is not straightforward due to the inhomogeneity of samples.
The disorder and surface roughness increase effectively the Gilbert
damping by adding a factor $\alpha_R$. Both $\alpha_R$ and
the nonlocal damping slow down the field-driven DW propagation.
\cite{Yuan2015_epjb, Yuan2015_prb} Weindler {\it et al.} measured
the Gilbert damping of collinear permalloy $\alpha=0.008$ and determined
$\alpha_R=0.003$ by comparing the experimental and simulated depinning
magnetic fields. By using the measured effective damping for the
field-driven DW motion $\alpha_{\rm eff}=0.023$, the nonlocal damping
parameter in the system of Ref. \cite{Weindler2014} is $\eta=1.3$~nm$^2$
that is denoted by the blue solid triangle in Fig.~\ref{fig3}.
We can also estimate $\eta$ by using the experimental data from Refs.~\onlinecite{Beach2005, Parkin2008}. The result is plotted in Fig.~\ref{fig3}
as a function of the Gilbert damping $\alpha+\alpha_R$. The inset of Fig.~\ref{fig3} is the experimental data (orange open squares) of the DW mobility
for different $\alpha$ obtained by Moore {\it et al.} \cite{Moore2010}.
They tuned the value of $\alpha$ by doping permalloy with a rare-earth element Holmium.
Micromagnetic simulation in the absence of the nonlocal damping results in
a significantly larger mobility (magenta empty triagles) than the measured
values. \cite{Moore2010} We use Eq.~(\ref{eq:vxvw}) to fit the experimental
data and perfectly reproduce the measured mobility as a function of $\alpha$.
Since we assume $\alpha_R=0$ in the fitting, the only fitted parameter
$\eta=5.1\pm1.6$~nm$^2$ corresponds to the upper bound of the nonlocal
damping strength.

The values of $\eta$ can also be extracted from the wave vector dependence of
the spin wave damping measured via FMR. \cite{Nembach2013,Schoen2015,Li2016}
The coexisting mechanisms such as the eddy current and the radiative
damping result in complexities in determining $\eta$. Nembach {\it et al.}
\cite{Nembach2013} found that $\eta=1.4$~nm$^2$ in permalloy nanodisks.
Li and Bailey \cite{Li2016} measured permalloy, cobalt and CoFeB
alloy and obtained the value $\eta=0.11\pm0.02$~nm$^2$ for permalloy.
Later Schoen{\it et al.}\cite{Schoen2015} discussed
the contribution of radiative damping and determined that
$\eta$ of permalloy from the experimental measurement is less
than 0.045~nm$^2$. All these experiments were performed at room
temperature while first-principles calculation at low temperature
found $\eta=0.016$~nm$^2$ in permalloy but it could be significantly
enhanced by two orders of magnitude, up to 5.9~nm$^2$, because
of the finite propagating length of transverse spin currents. \cite{Yuan2014}

%%%%%%%%%%%%%%%%%%%%%%%%%%%%%%%%%%%%%%%%%%%%%%%%%%%%%%%%%%%%%%
In the derivation of the GTE~(\ref{thiele}), we have included
both the transverse and longitudinal torques [Eq.~(\ref{eq:tsc}) and Eq.~(\ref{eq:lsc})] due to the spin pumping.
For the steady-state DW motion described by Eq.~(\ref{steady}),
the longitudinal component $\bm{\tau}_{\rm LSC}=-\eta\partial_i(
\partial_i\mathbf m \times \partial_t\mathbf m)$ vanishes because
$\partial_i\mathbf m$ is aligned with $\partial_t\mathbf m$.
In our numerical simulation for transverse DWs, it is confirmed that
$\bm{\tau}_{\rm LSC}$ alone did not change the DW velocity below
breakdown field, consistent with conclusion in Ref.~\onlinecite{Zhang2009}
where the expression of $\bm{\tau}_{\rm LSC}$ was derived.

In contrast, Weindler {\it et al.} \cite{Weindler2014} considered
only $\bm{\tau}_{\rm LSC}$ in their micromagnetic simulations with
$\eta=0.07$~nm$^2$ and found the texture-enhanced damping in the
steady-state DW motion. In micromagnetic simulations, there might be
higher order effects that eventually affect the DW mobility.
On the other hand, $\bm{\tau}_{\rm TSC}$ can have remarkable influence
on the field-driven DW velocity, as predicted by Eq.~(\ref{eq:vxvw}).
This is partly the reason why we extracted a different $\eta=1.3$~
nm$^2$ by using the same experimental data in Sec.~\ref{sec:vdw}.
It suggests that $\bm{\tau}_{\rm TSC}$ must be properly included
to extract a reliable value of $\eta$.

\section{Conclusions\label{sec:IV}}
We have derived the nonlocal damping torque originated from the pumped spin current by the precessional noncollinear magnetization.
This nonlocal damping torque consists of a longitudinal and a transverse components that both depend on the magnetization gradient and inevitably affects the noncollinear magnetization dynamics. We derive a generalized Thiele equation (\ref{thiele}) for the field-driven steady-state DW motion under the influence of the nonlocal damping.
For both transverse and vortex DWs in ferromagnetic nanowires, the transverse component of the nonlocal
damping slows down the DW propagation and increases the Walker breakdown field.
The analytical results are further confirmed by numerical simulations for
transverse DWs. In addition, our result compares well the experimentally
reported damping dependence of vortex DW mobility under an external magnetic
field, while the LLG equation without the
nonlocal damping significantly overestimates the DW velocity.

The nonadiabatic spin transfer torque $\beta$ is one of the key parameters in current-driven DW motion because the velocity is proportional to $\beta$ and inversely proportional to the damping 
$\alpha$. Previous works in literature extracted $\beta$ value from experimentally 
measured DW velocity by assuming a constant $\alpha$ in a DW. 
Thus the influence of the nonlocal damping revealed in this paper may explain the 
large spread in the extracted $\beta$ values  for permalloy. \cite{Hayashi2008,Lepadatu2009,Eltschka2010,Chauleau2014} The measurement of current-induced torques due to Rashba and Dzyaloshinskii-Moriya interactions in noncollinear magnetization is usually performed by comparing the experimental observation and micromagnetic simulation. As a prerequisite, the nonlocal damping has to be appropriately included in the simulations while the damping form may be more complicated because of the complex interactions.\cite{Kim2015} 

%%%%%%%%%%%%%%%%%%%%%%%%%%%%%%%%%%%%%%%%%%%%%%%%%%%%%%%%%%%%%%
\appendix
\section{Spin pumping and the damping torque in a noncollinear ferromagnetic
nanowire}\label{app:mixing}

\begin{figure}[b]
\centering
\includegraphics[width=\columnwidth]{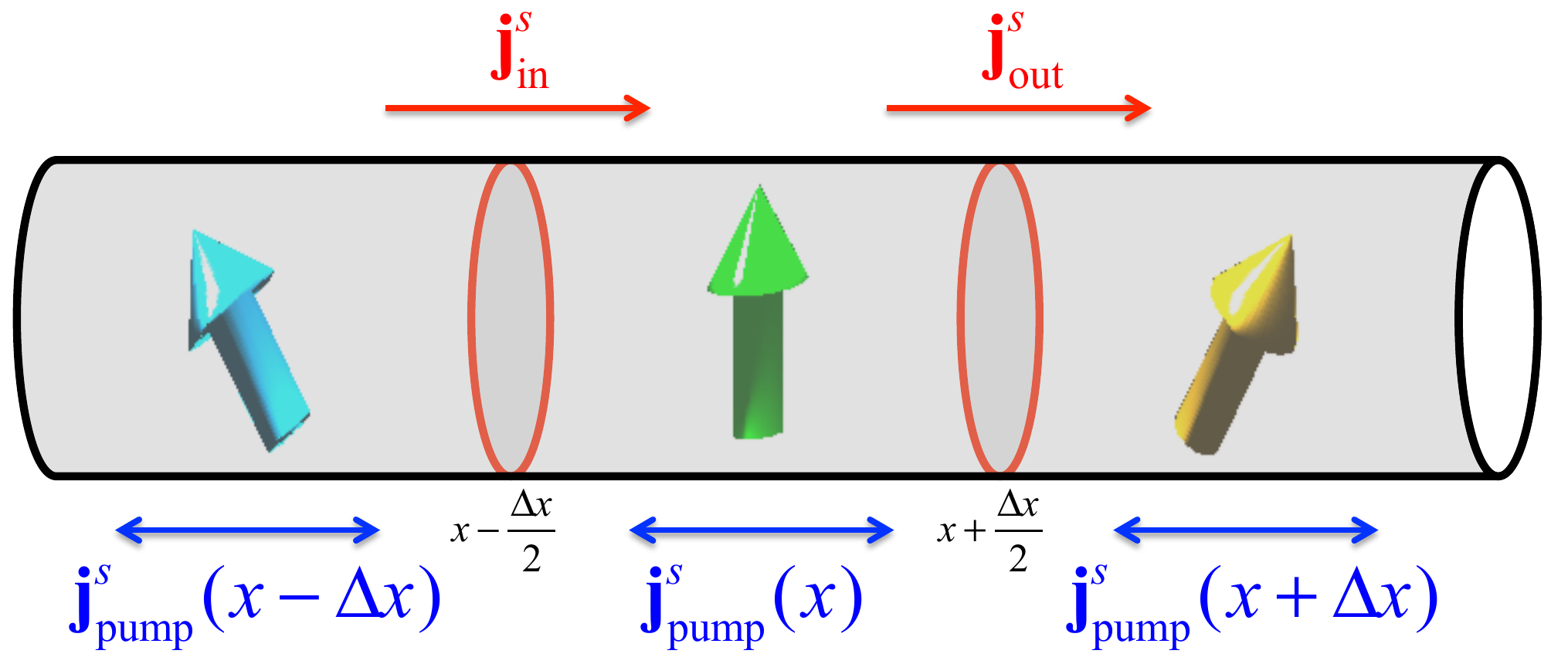}
\caption{Sketch of spin pumping in a ferromagnetic nanowire with
noncollinear magnetization. A segment of magnetization pumps a spin
current $\mathbf j^s_{\rm pump}(x)=\frac{\hbar}{4\pi}g_{\uparrow\downarrow}
\mathbf m(x)\times\partial_t\mathbf m(x)$. The net spin current
flowing into (out of) the segment at $x$ is given by the backward
(forward) derivative of the pumped spin current. The torque exerted
on the segment of magnetization at $x$ is then defined as the net
spin current that is absorbed by the segment of magnetization, as
formulated by Eq.~(\ref{eq:abs}).}
\label{fig4}
\end{figure}
In this appendix, we derive spin pumping Eq.~(\ref{eq:is}) and the resulting
damping torque Eq.~(\ref{eq:damping}) exerted on the local magnetization in a
noncollinear ferromagnetic nanowire.
Without loss of generality, we consider a one-dimensional case with
the magnetization $\mathbf m(x,t)$, in which a small segment of
magnetization centered at position $x$ can be approximated by
a local collinear magnetization $\mathbf m(x)$. As $\mathbf m(x)$
varies in time, it can pump out a spin current
\begin{equation}
\mathbf j_{\rm pump}^s(x)=\frac{\hbar}{4\pi}g_{\uparrow\downarrow}
\mathbf m(x)\times\partial_t\mathbf m(x),
\end{equation}
where $g_{\uparrow\downarrow}$ is the conventional spin-mixing conductance characterizing
the magnitude of spin pumping. \cite{Tserkovnyak2002} This pumped spin current flows
both forward and backward, as schematically shown by the blue arrows in Fig.~\ref{fig4}.
Since the magnetization in the nanowire is not uniform, the spin current pumped forward
by the segment at position $x$ is different from that pumped backward by the next segment
at position $x+\Delta x$. The incomplete cancellation of these two pumped spin currents
gives rise to a net spin currents across the cross section at $x+\Delta x/2$, which reads
\begin{eqnarray}
\mathbf j^s\left(x+\frac{\Delta x}{2}\right)&=&\mathbf j_{\rm pump}^s(x)-\mathbf j_{\rm pump}^s(x+\Delta x)\nonumber\\
&=&\frac{\hbar}{4\pi}g_{\uparrow\downarrow}\left[\mathbf m(x)\times
\partial_t\mathbf m(x)\right.\nonumber\\&&\left.-\mathbf m(x+\Delta x)\times\partial_t\mathbf m(x+\Delta x)
\right]\nonumber\\
&=&-\frac{\hbar}{4\pi}g_{\uparrow\downarrow}\Delta x\,\partial_x\left[\mathbf m(x)\times
\partial_t\mathbf m(x)\right]_{x+\frac{\Delta x}{2}}.\nonumber\\\label{eq:spinout}
\end{eqnarray}
If we choose $+x$ as the positive direction, $\mathbf j^s(x+\Delta x/2)$ corresponds to
the outflow of the spin current for the segment of magnetization at $x$, which will later
be referred to as $\mathbf j^s_{\rm out}$. $\mathbf j^s_{\rm out}$ can be rewritten as
a superposition of a longitudinal component $\mathbf j^{s\parallel}_{\rm out}$ that
is aligned with $\mathbf m(x)$ and a transverse one $\mathbf j^{s\perp}_{\rm out}$
that is perpendicular to $\mathbf m(x)$,
\begin{eqnarray}
\mathbf j^{s\parallel}_{\rm out}&=&-\frac{\hbar}{4\pi}g_{\uparrow\downarrow}\Delta x\left[\partial_x\mathbf m(x)\times
\partial_t\mathbf m(x)\right]_{x+\frac{\Delta x}{2}},\\
\mathbf j^{s\perp}_{\rm out}&=&-\frac{\hbar}{4\pi}g_{\uparrow\downarrow}\Delta x\left[\mathbf m(x)\times
\partial_x\partial_t\mathbf m(x)\right]_{x+\frac{\Delta x}{2}}.
\end{eqnarray}
In the same manner, we are able to find the the net spin current across the
cross section at $x-\Delta x/2$ corresponding to the inflow,
\begin{eqnarray}
\mathbf j^s_{\rm in}&\equiv&\mathbf j^s\left(x-\frac{\Delta x}{2}\right)
=\mathbf j_{\rm pump}^s(x-\Delta x)-\mathbf j_{\rm pump}^s(x)\nonumber\\
&=&-\frac{\hbar}{4\pi}g_{\uparrow\downarrow}\Delta x\,\partial_x\left[\mathbf m(x)\times
\partial_t\mathbf m(x)\right]_{x-\frac{\Delta x}{2}}.\label{eq:spinin}
\end{eqnarray}
Its longitudinal and transverse components can be respectively written as
\begin{eqnarray}
\mathbf j^{s\parallel}_{\rm in}&=&-\frac{\hbar}{4\pi}g_{\uparrow\downarrow}\Delta x\left[\partial_x\mathbf m(x)\times
\partial_t\mathbf m(x)\right]_{x-\frac{\Delta x}{2}},\\
\mathbf j^{s\perp}_{\rm in}&=&-\frac{\hbar}{4\pi}g_{\uparrow\downarrow}\Delta x\left[\mathbf m(x)\times
\partial_x\partial_t\mathbf m(x)\right]_{x-\frac{\Delta x}{2}}.
\end{eqnarray}

The difference between inflow and outflow spin currents is absorbed by the local magnetization
resulting a damping torque
\begin{eqnarray}
\bm{\tau}_{\rm damping}=-\frac{\gamma A\left(\mathbf j^s_{\rm in}
-\mathbf j^s_{\rm out}\right)}{M_s A \Delta x}=\frac{\gamma}{M_s}\partial_x\mathbf j^s(x).\label{eq:abs}
\end{eqnarray}
where $A$ is the cross sectional area of the nanowire and the prefactor
$\gamma/(M_s A\Delta x)$ is to convert the torque in the unit of s$^{-1}$.
The torque in Eq.~(\ref{eq:abs}) due to absorption of pumped spin current
is only determined by the local magnetization gradient, implying that the
length scale of magnetization variation, e.g. a DW width or a spin wave length,
is much larger than the propagating length of the pumped spin current
in the ferromagnetic nanowire. This condition is in general satisfied
in disordered ferromagnetic materials, but it is found in permalloy at low temperature
that the spin coherent length is up to 13.1~nm, where the nonlocal
absorption of the pumped spin current even changes the scaling of the
effective damping with the DW width. \cite{Yuan2014}

While Eq.~(\ref{eq:spinout}) and Eq.~(\ref{eq:spinin}) are two particular
examples, the net spin current in a noncollinear magnetization resulting
from spin pumping can be generally expressed as
\begin{eqnarray}
\mathbf j^s(x,t)&=&-\frac{\hbar}{4\pi}g_{\uparrow\downarrow} \Delta x\,
\partial_x \left[\mathbf m(x,t)\times\partial_t\mathbf m(x,t)\right]\nonumber\\
&=&-\frac{\hbar}{4\pi}\Gamma_{\uparrow\downarrow} \,\partial_x
\left[\mathbf m(x,t)\times\partial_t\mathbf m(x,t)\right].\label{eq:spinnet}
\end{eqnarray}
Here we define the spin-mixing conductivity $\Gamma_{\uparrow\downarrow}
\equiv g_{\uparrow\downarrow}\Delta x$ that is a proper parameter
characterizing the (intralayer) spin pumping in a noncollinear ferromagnetic material.
The conventional spin pumping at a ferromagnet-normal metal interface
\cite{Tserkovnyak2002} can be recovered from Eq.~(\ref{eq:spinnet})
by taking the effective thickness of the interface $\Delta x$.
Notice that $g_{\uparrow\downarrow}$ has the dimension of the inverse
of area and $\Gamma_{\uparrow\downarrow}$ has the dimension of the inverse
of length.

\acknowledgments
H. Y. Y. would like to thank Lei Wang for helpful discussions.
This work was financially supported by National Basic Research Program of China
(Grant No. 2012CB921304), National Natural Science Foundation of China (Grant No. 61376105).
X. R. W. was supported by the National Natural Science
Foundation of China (Grant No. 11374249) as well as Hong
Kong RGC Grants No. 163011151 and No. 605413.

\end{document}